\documentclass[aps,prl,twocolumn,showpacs,showkeys,amsmath,amssymb]{revtex4}
\usepackage{amsfonts}
\usepackage{amsmath}
\usepackage{graphicx}
\usepackage{subfigure}
\usepackage{dcolumn}
\usepackage{bm}
\usepackage{booktabs}
\usepackage[utf8]{inputenc}

\usepackage{graphicx,graphics,dcolumn,booktabs,bm}
\usepackage{longtable,lscape}
\usepackage{txfonts}
\usepackage{overpic}
\usepackage{amssymb}
\usepackage{indentfirst}
\usepackage{epsfig}
\usepackage{feynmf}   
\usepackage{epstopdf}   
\usepackage{slashed}  
\usepackage{multirow}

\usepackage[colorlinks, citecolor=blue,anchorcolor=red,menucolor=red, linkcolor=red,filecolor=red,runcolor=red,urlcolor=blue,frenchlinks=red]{hyperref}

\usepackage{ulem}
\usepackage[usenames,dvipsnames]{color}

\makeatletter
\newcommand{\figcaption}{\def\@captype{figure}\caption}
\newcommand{\tabcaption}{\def\@captype{table}\caption}

\newcommand{\Rmnum}[1]{\expandafter\@slowromancap\romannumeral #1@}
\def\hlinewd#1{%
  \noalign{\ifnum0=`}\fi\hrule \@height #1 \futurelet
   \reserved@a\@xhline}
\makeatother

\newcommand\dqq{\left< \bar s s \right>}
\newcommand\dqGq{\left< g_s \bar s \sigma G s \right>}
\newcommand\dGG{\left< g_s^2 GG \right>}
\newcommand\unit{\mathrm{GeV}}

\newcommand\Msq{M_B^2}
\newcommand\M{M_B}

\newcommand\mycite{\cite}

\begin{document}


\title{Exotic $\Omega\Omega$ dibaryon states in a molecular picture}

\author{Xiao-Hui Chen$^1$}
\author{Qi-Nan Wang$^1$}
\author{Wei Chen$^1$}
\email{chenwei29@mail.sysu.edu.cn}
\author{Hua-Xing Chen$^2$}
\email{hxchen@buaa.edu.cn}
\affiliation{$^1$School of Physics, Sun Yat-Sen University, Guangzhou 510275, China
\\
 $^2$School of Physics and
Beijing Key Laboratory of Advanced Nuclear Materials and Physics, \\
Beihang University, Beijing 100191, China }

\begin{abstract}
We investigate the exotic $\Omega\Omega$ dibaryon states with $J^P=0^+$ and $2^+$ in a molecular picture. We construct the scalar and tensor  $\Omega$$\Omega$ molecular interpolating currents and calculate their masses within the method of QCD sum rules. Our results indicate that the mass of the scalar dibaryon state is $m_{\Omega\Omega, \, 0^+}=(3.33\pm0.22) \,\unit$, which is about $15 \,\mathrm{MeV}$ below the $2m_\Omega$ threshold. This result suggests the existence of a loosely bound molecular state of the $J^P=0^+$ scalar $\Omega\Omega$ dibaryon with a small binding energy around 15 MeV. The mass of the tensor dibaryon is predicted to be $m_{\Omega\Omega,\, 2^+}=(3.24\pm0.23)\, \mbox{GeV}$, which may imply a deeper molecular state of the tensor $\Omega\Omega$ dibaryon than the scalar channel. These exotic strangeness $S=-6$ and doubly-charged $\Omega\Omega$ dibaryon states may be identified in the heavy-ion collision processes. 
\end{abstract}


\maketitle

\noindent {\it Introduction:} 
The history of multiquark configurations can go back to the establishment of the quark model (QM) by Gell-Mann~\cite{1964-Gell-Mann-p214-215} and Zweig~\cite{1964-Zweig-p-}, where the tetraquark $qq\bar q\bar q$ and pentaquark $qqqq\bar q$ configurations were proposed outside of the conventional meson and baryon states. In the past seventeen years, there have been great progress on the explorations of tetraquark and pentaquark states, with the observations of the so called $XYZ$ and $P_c$ states~\cite{Liu:2019zoy,2016-Chen-p1-121,2016-Esposito-p1-97,2016-Richard-p1185-1212,2017-Ali-p123-198,2017-Lebed-p143-194,2018-Guo-p15004-15004,2018-Olsen-p15003-15003}.

A dibaryon is another kind of multiquark system composed of two color-singlet baryons, such as the deuteron (a loosely $np$ bound state in $^3S_1$ channel~\cite{deuteron}). In 1964, the non-strange dibaryon sextet $D_{IJ}$ (with $IJ=01,10,12,21,03$ and $30$) was proposed by Dyson and Xuong in SU(6) symmetry~\mycite{dibaryon-sextet}. The $D_{01}$, $D_{10}$ and $D_{12}$ dibaryons have been identified as the deuteron ground state, the virtual $^1S_0$ isovector state and an isovector $J^P=2^+$ state at the $\Delta N$ threshold respectively~\mycite{dibaryon-sextet}. Recently, the $d^* (2380)$ state was confirmed by the WASA detector at COSY~\cite{2011-Adlarson-p242302-242302,2013-Adlarson-p229-236,2014-Adlarson-p202301-202301,2015-Adlarson-p325-332}, which was considered as the $\Delta\Delta$ dibaryon in the $D_{03}$ channel~\cite{1999-Yuan-p45203-45203,2013-Gal-p172301-172301,2015-Chen-p25204-25204,2014-Huang-p34001-34001,2015-Huang-p71001-71001}. 
Moreover, the H-dibaryon predicted by Jaffe~\mycite{H-dibaryon} is still attractive both in experimental and theoretical aspects~\mycite{H-dibaryon-experiment1,H-dibaryon-theoretical1,H-dibaryon-theoretical2}. For more introduction about dibaryons, one can consult the recent review paper in Ref.~\mycite{dibaryon-review}.
	
Comparing to the $NN$ and H dibaryons, the investigation on the $\Omega\Omega$ system has received much less research interest. 
The interaction between two $\Omega$ baryons has not been adequately understood experimentally and theoretically. Nevertheless, one would expect that the $\Omega\Omega$ dibaryon will be stable against the strong interaction, since $\Omega$ is the only stable state in the decuplet $\bf{10}$ baryons~\cite{2018-Tanabashi-p30001-30001}. From the properties of $\Omega$, we know the baryon number of $\Omega\Omega$ is $2$ and the stangeness $S=-6$, which is the most strange dibaryon state. Under the restriction of Pauli exclusion principle, the total wave function of the $\Omega\Omega$ system should be antisymmetric, which results in the even total spin $S=0$ or $S=2$ for the S-wave ($L=0$) coupling. The spin-parity quantum numbers is thus $J^P=0^+$ or $2^+$ and there is no isospin for such a system. 

To date, the $\Omega\Omega$ dibaryon states have been studied in a quark potential model~\cite{1995-Wang-p3411-3411}, the chiral SU(3) quark model~\cite{2000-Zhang-p65204-65204}, and the lattice QCD simulations~\cite{2012-Buchoff-p94511-94511,2018-Gongyo-p212001-212001}. In a quark potential model of Ref.~\cite{1995-Wang-p3411-3411}, the authors calculated the effective interaction between two $\Omega$ baryons by including the quark delocalization and color screening. They found that the mass of the scalar $\Omega\Omega$ system was heavier than the $2m_\Omega$ threshold, resulting in a weekly repulsive interaction. This result was supported by the lattice QCD calculation at a pion mass of 390 MeV in Ref.~\cite{2012-Buchoff-p94511-94511}, where the weakly repulsive interactions were found for both the $S=0$ and $S=2$ $\Omega\Omega$ systems. In the chiral SU(3) quark model, the structure of the scalar $\Omega\Omega$ dibaryon was studied by solving a resonating group method equation~\cite{2000-Zhang-p65204-65204}. Their result suggested a deep attraction with binding energy around 100 MeV. In Ref.~\cite{2015-Yamada-p1-71}, the HAL QCD Collaboration investigated the interaction between two $\Omega$ baryons at $m_\pi=700$ MeV and found that the $\Omega\Omega$ potential has a repulsive core at short distance and an attractive well at intermediate distance. The phase shift obtained from the potential showed moderate attraction at low energies. 
Recently, the HAL QCD Collaboration has performed the $(2+1)$-flavor lattice QCD simulations on the $(\Omega\Omega)_{0^+}$ dibaryon at nearly physical pion mass $m_\pi=146$ MeV~\cite{2018-Gongyo-p212001-212001}. They found an overall attraction for the scalar $\Omega\Omega$ dibaryon with a small binding energy $B_{\Omega\Omega}=1.6$ MeV. These conflicting results from the phenomenological models and lattice simulations are inspiring more theoretical studies for the $\Omega\Omega$ dibaryon systems. In this work, we shall study the $\Omega\Omega$ dibaryons in both the scalar $^1S_0$ and tensor $^5S_2$ channels in the QCD sum rule method. 
	

\noindent {\it QCD Sum Rules for Dibaryon Systems:} 
In the past several decades, the QCD sum rule has been used as a powerful non-perturbative approach to investigate the hadron properties, such as the hadron masses, magnetic moments, decay widths and so on \mycite{QCD-sum-rule-shifman, QCD-sum-rule-Reinders}. To study the dibaryon systems in QCD sum rules, we need to construct the $\Omega\Omega$ interpolating currents by using 
the local Ioffe current for the $\Omega$ baryon~\cite{1981-Ioffe-p317-341,1983-Ioffe-p67-67}
\begin{equation}
J_\mu^\Omega(x)=\epsilon^{abc}\left[s^T_a (x) C\gamma_\mu s_b (x) \right]s_c (x)\, ,
\end{equation}
in which $s(x)$ represents the strange quark field, $a, b, c$ are the color indices, $\gamma_\mu$ is the Dirac matrix,  $C=i\gamma_2\gamma_0$ is the charge conjugation matrix and $T$ the transpose operator. The $\Omega\Omega$ dibaryon interpolating 
currents are then composed in the molecular picture as 
\begin{eqnarray}
\nonumber J^{\Omega\Omega}(x) &=&\epsilon^{abc}\epsilon^{def} \left[s^T_a (x) C\gamma_\mu s_b (x) \right]s^T_c (x) 
\cdot C\gamma_5 \cdot 
\\ &&
s_f (x) \left[s^T_d (x) C\gamma^\mu s_e (x) \right]  \label{scalarcurrent}
\end{eqnarray}
with $J^P=0^+$, and 
\begin{eqnarray}
\nonumber J^{\Omega\Omega}_{\mu\nu}(x) &=&\epsilon^{abc}\epsilon^{def} \left[s^T_a (x) C\gamma_\mu s_b (x) \right]s^T_c (x) 
\cdot C\gamma_5 \cdot 
\\ &&
s_f (x) \left[s^T_d (x) C\gamma_\nu s_e (x) \right] \label{tensorcurrent}
\end{eqnarray}
with $J^P=2^+$. 
With these interpolating currents, we consider the two-point correlation functions for $\Omega\Omega$ dibaryon systems 
\begin{eqnarray} 
\Pi (q^2) &=& i \int\mathrm{d^4} x \  \mathrm{e}^{iq\cdot x} \left< 0 \left| \mathrm{T} \left\{ J^{\Omega\Omega} (x)  J^{\Omega\Omega\dag} (0) \right\} \right| 0 \right>\, ,  \label{scalarCF}
\\
\Pi_{\mu\nu,\,\rho\sigma}(q^2) &=& i \int\mathrm{d^4} x \  \mathrm{e}^{iq\cdot x} \left< 0 \left| \mathrm{T} \left\{ J^{\Omega\Omega}_{\mu\nu}(x)  J^{\Omega\Omega\dag}_{\rho\sigma}(0) \right\} \right| 0 \right>\, ,  \label{tensorCF}
\end{eqnarray}
where $J^{\Omega\Omega}(x)$ and $J^{\Omega\Omega}_{\mu\nu}(x)$ can respectively couple to the scalar and tensor dibaryon states we interested in 
\begin{align}
\langle0|J^{\Omega\Omega}|X\rangle&=f_{S}\,, \label{scalarcoupling}\\
\langle0|J^{\Omega\Omega}_{\mu\nu}|X\rangle&=f_{T} \epsilon_{\mu\nu}+\cdots\, ,
\label{tensorcoupling}
\end{align}
in which $f_S$ and $f_T$ are the coupling constants, $\epsilon_{\mu\nu}$ is the polarization tensor 
coupling to the spin-2 state. Noting that $J^{\Omega\Omega}_{\mu\nu}(x)$ is a symmetric operator, it can induce both the $J^P=0^+$ and $2^+$ invariant functions with the following projectors~\cite{2014-Chen-p201-215,2017-Chen-p114005-114005,2017-Chen-p114017-114017}
\begin{equation}
\begin{split}
P_{0T}&=\frac{1}{16}g_{\mu\nu}g_{\rho\sigma}\, , ~~~~~~~\mbox{for}\, J^P=0^+,\, \mbox{T}\\
P_{0S}&=T_{\mu\nu}T_{\rho\sigma}\, , ~~~~~~~~~~\mbox{for}\, J^P=0^+,\, \mbox{S}\\
P_{0TS}&=\frac{1}{4}(T_{\mu\nu}g_{\rho\sigma}+T_{\rho\sigma}g_{\mu\nu})\, , ~\mbox{for}\, J^P=0^+,\, \mbox{TS}\\
P_{2S}^P&=\frac{1}{2}\left(\eta_{\mu\rho}\eta_{\nu\sigma}+\eta_{\mu\sigma}\eta_{\nu\rho}-\frac{2}{3}\eta_{\mu\nu}\eta_{\rho\sigma}\right)\, , ~\mbox{for}\, J^P=2^+,\, \mbox{S} \label{projectors}
\end{split}
\end{equation}
where
\begin{equation}
\begin{split}
\eta_{\mu\nu}&=\frac{q_\mu q_\nu}{q^2}-g_{\mu\nu}\, ,~~~
T_{\mu\nu}=\frac{q_\mu q_\nu}{q^2}-\frac{1}{4}g_{\mu\nu}\, , \\
T_{\mu\nu,\rho\sigma}^\pm&=\left[\frac{q_\mu q_\rho}{q^2}\eta_{\nu\sigma}\pm(\mu\leftrightarrow\nu)\right]\pm(\rho\leftrightarrow\sigma)\, .
\end{split}
\end{equation}
The projectors $P_{0T}$, $P_{0S}$ and $P_{0TS}$ in Eq.~\eqref{projectors} can be used to pick out different invariant functions induced by the the trace part (T), traceless symmetric part (S) and their cross term part (TS) from the tensor current respectively, which all couple to the $J^P=0^+$ channel with different coupling constants. 

At the hadronic level, the invariant structure of correlation function $\Pi(q^2)$ can be expressed as a dispersion relation 
\begin{equation}
\Pi\left(q^{2}\right)=\left(q^{2}\right)^{N}\int_{0}^{\infty}\mathrm{d}s\frac{\rho\left(s\right)}{s^{N}\left(s-q^{2}-\mathrm{i}\epsilon\right)}+\sum_{k=0}^{N-1}b_{n}\left(q^{2}\right)^{k}\,
, \label{eq:dispersion relation}
\end{equation}
where $b_n$ is an unknown subtraction constant. The spectral function can be usually written as a sum over 
$\delta$ functions by inserting intermediate states $|n\rangle$ with the same quantum numbers as the interpolating current
\begin{align}
\rho\left(s\right) &\equiv\mathrm{Im}\Pi\left(s\right)/\pi =f_{X}^{2}\delta\left(s-m_{X}^{2}\right)\langle 0|J |n\rangle \langle n |J^{\dagger} |0 \rangle \nonumber \\
 & =f_{X}^{2}\delta\left(s-m_{X}^{2}\right)+\text{continuum}\, ,
\end{align}
where we adopt the ``narrow resonance" approximation to describe the spectral function, and $m_{X}$ is the mass of the lowest-lying resonance $X$.

The correlation functions can also be calculated as the functions of various QCD condensates at the quark-gluonic level, using the operator product expansion(OPE) method. These results shall be equal to the correlation function in Eq.~\eqref{eq:dispersion relation} via the quark-hadron duality, after performing the Borel transform to remove the unknown subtraction constants and suppress the continuum contributions
\begin{eqnarray}
\Pi(s_0,\, M_B^2)=f_X^2m_X^{2}e^{-m_X^2/M_B^2}=\int_{<}^{s_0}dse^{-s/M_B^2}\rho(s)\, ,
\label{sumrule}
\end{eqnarray} 
in which $s_{0}$ is the continuum threshold and $M_{B}$ is the Borel mass. 
Then we can calculate the hadron mass as 
\begin{equation}
m^{2}_X\left(s_0,\, M_B^2\right)=\frac{\int_{<}^{s_{0}}ds\,s\rho\left(s\right)e^{-s/M_{B}^{2}}}{\int_{<}^{s_{0}}ds\,\rho\left(s\right)e^{-s/M_{B}^{2}}}\, , \label{hadronmass}
\end{equation}
in which the spectral density $\rho(s)$ is evaluated in the quark-gluonic level as the function of various QCD condensates up to dimension-10, including the quark condensate $\langle \bar ss\rangle$, quark-gluon mixed condensate $\langle g_s\bar s\sigma\cdot Gs\rangle$, gluon condensate $\langle g_s^2GG\rangle$ and so on. For the dibaryon currents $J^{\Omega\Omega}(x)$ and $J^{\Omega\Omega}_{\mu\nu}(x)$, the expressions of spectral densities are lengthy so that we will show them in the Supplemental Material. 

{\it Prediction for the scalar $\Omega\Omega$ Dibaryon with $J^P=0^+$:} We use the following values for various QCD parameters in our numerical analyses~\mycite{PDG, condensates1, condensates2, condensates3, condensates4, condensates5, condensates-chenhuaxing}

\begin{equation}
	\begin{array}{cc}
		\toprule
		\dqq & - (0.8 \pm 0.1) \times (0.24\pm0.03)^3 \, \unit^3 \\
		\dGG & (0.48 \pm 0.14) \, \unit^4 \\
		\dqGq & - M_0^2 \dqq \\
		M_0^2 & (0.8 \pm 0.2) \, \unit^2 \\
		m_s & 95 ^{+9} _{-3} \, \mathrm{MeV} \\
		\bottomrule
	\end{array} \label{QCDparameters}
\end{equation}

\begin{figure}[ht]
	\includegraphics[width=0.47\textwidth]{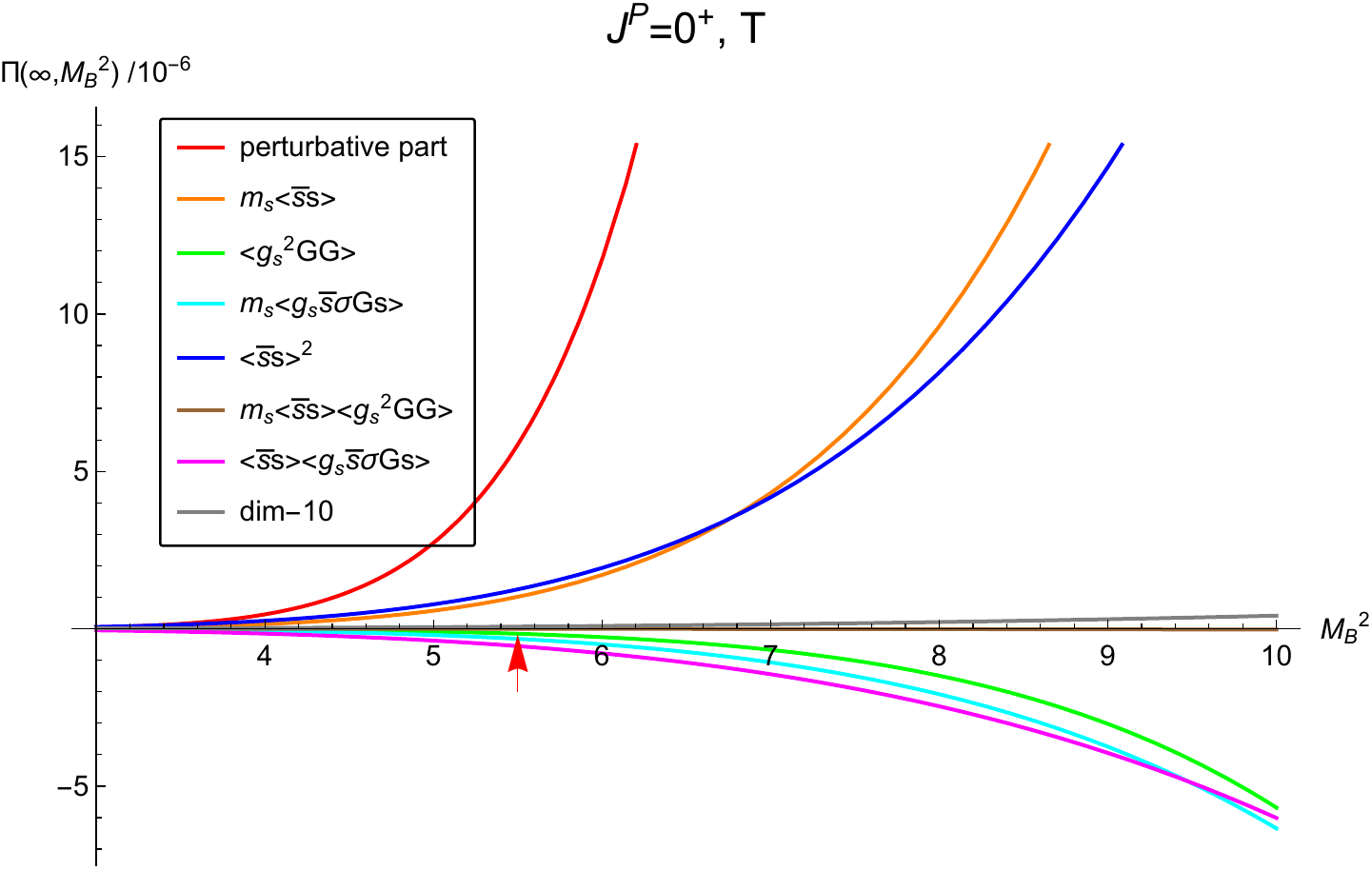}
	\caption{OPE convergence for the trace part (T) in $J^{\Omega\Omega}_{\mu\nu}(x)$.}
	\label{fig:ope0T}
\end{figure}

\begin{figure}[ht]
	\includegraphics[width=0.43\textwidth]{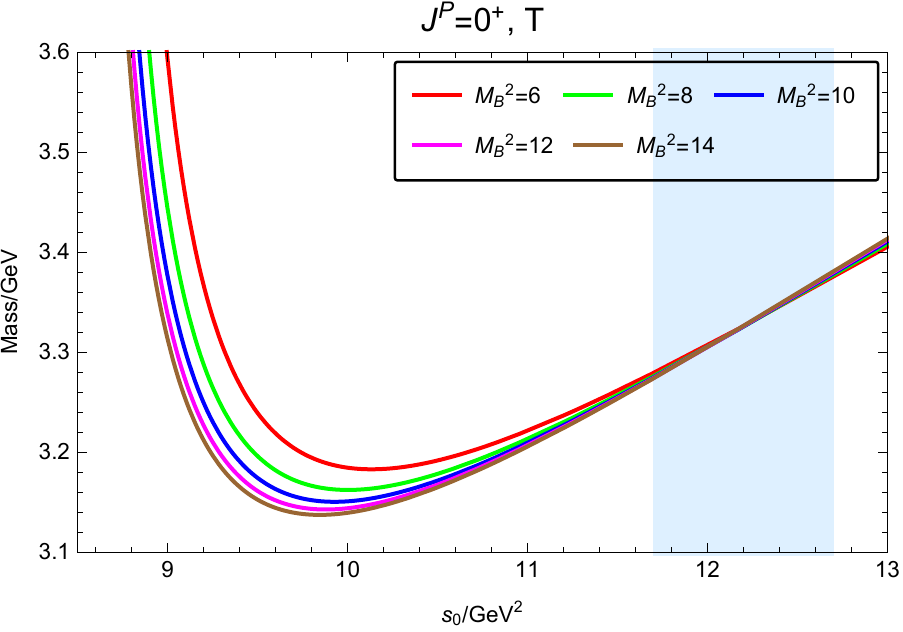}
	\quad
	\includegraphics[width=0.43\textwidth]{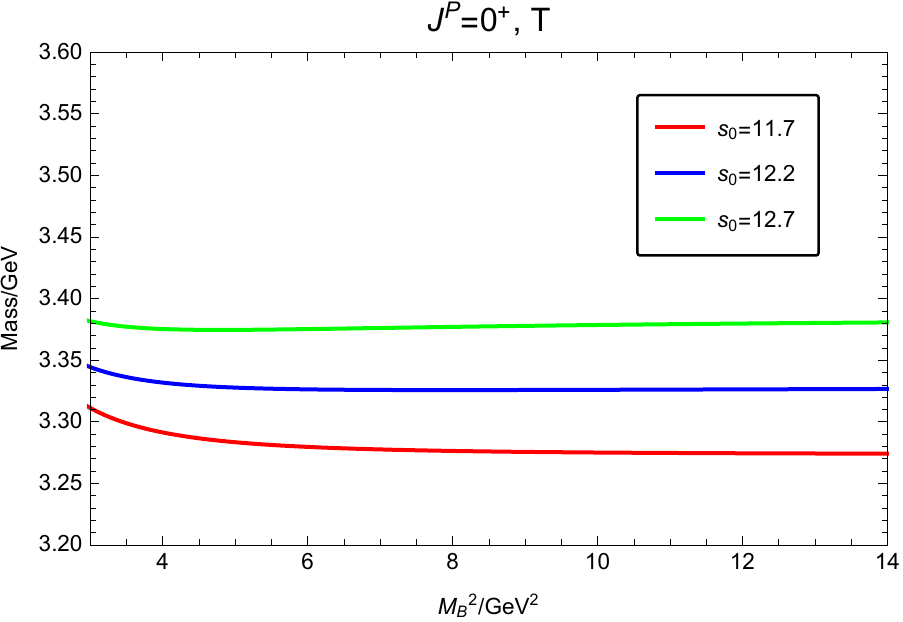}
	\caption{Extracted hadron mass for the trace part (T) in $J^{\Omega\Omega}_{\mu\nu}(x)$.}
	\label{fig: result tensor 0 T}
\end{figure}

In Eq.~\eqref{hadronmass}, the hadron mass is extracted as a function of two free parameters: the Borel mass $M_B$ and the continuum threshold $s_0$. For the numerical analysis, we study the OPE convergence to determine the lower bound on the Borel mass $M_B$, requiring the quark condensate and the four-quark condensate be less than one third of the perturbative term while the dimension-10 condensate contribute less than $1\%$. For the tensor current $J^{\Omega\Omega}_{\mu\nu}(x)$, we show the two-point correlation function numerically of the trace part (T) coupling 
to $J^P=0^+$ 
\begin{align}
& \Pi (\infty,\, \Msq) = 6.98 \times 10^{-12} \M^{16} + 3.22 \times 10^{-11} \M^{12} 
\\ \nonumber
&+ 1.83 \times 10^{-10} \M^{10}- 6.04 \times 10^{-10} \M^8 + 4.21 \times 10^{-10} \M^6\, , 
\end{align}
in which we take $s_0\to\infty$. We plot the correlation function $\Pi (\infty,\, \Msq)$ term by term in Fig.~\ref{fig:ope0T}, in which the quark condensate $m_s\dqq$ and four-quark condensate $\dqq^2$ are much larger than any other non-perturbative term. Accordingly, the lower bound on the Borel mass can be obtained as $M_B^2\geq 5.5$ GeV$^2$. In principle, the upper bound on the Borel mass can be obtained by studying the pole contribution. However, the high power behavior ($s^7$ for the perturbative term) in the dibaryon spectral density leads to small pole contribution and thus fails to constrain the upper bound on the Borel mass~\mycite{QCD-sum-rule-chenhuaxing}. Instead, we shall require good stability of the Borel curves. For the continuum threshold parameter, an optimized choice of $s_0$ is the value minimizing the variation of the hadron mass with the Borel mass. As shown in Fig.~\ref{fig: result tensor 0 T}, the working region of $s_0$ can be obtained as 11.7 GeV$^2$ $\leq s_0\leq 12.7$ GeV$^2$. 
%
%

Within these parameter values, we plot the Borel curves of the extracted hadron mass in Fig.~\ref{fig: result tensor 0 T}. 
These Borel curves behave very good stability and give mass prediction of the scalar 
$\Omega\Omega$ dibaryon with $J^P=0^+$ (T) as 
\begin{align}
m_{\Omega\Omega}=(3.33\pm0.14)\, \mbox{GeV}\, , \label{massTtensor}
\end{align}
in which the errors come from the uncertainties of $M_B$, $s_0$ and various QCD parameters in Eq.~\eqref{QCDparameters}. The corresponding coupling constant can also be evaluated as 
\begin{align}
f_{\Omega\Omega}=\left( 4.6\pm0.7 \right) \times 10^{-4}\, \mbox{GeV}^8\, .
\end{align}

As indicated in Eq.~\eqref{projectors}, the traceless symmetric part (S) and cross term (TS) in the tensor correlation function $\Pi_{\mu\nu,\,\rho\sigma}(q^2)$ can also couple to the scalar $\Omega\Omega$ channel with $J^P=0^+$. 
However, our analyses indicate that the dibaryon sum rule from the traceless symmetric part (S) is unstable against the Borel mass, which cannot provide reliable mass prediction. For the case of TS cross term, the perturbative term in the OPE series is absent, and thus we will not use this invariant structure to study the scalar dibaryon. 

\begin{figure}[ht]
\includegraphics[width=0.43\textwidth]{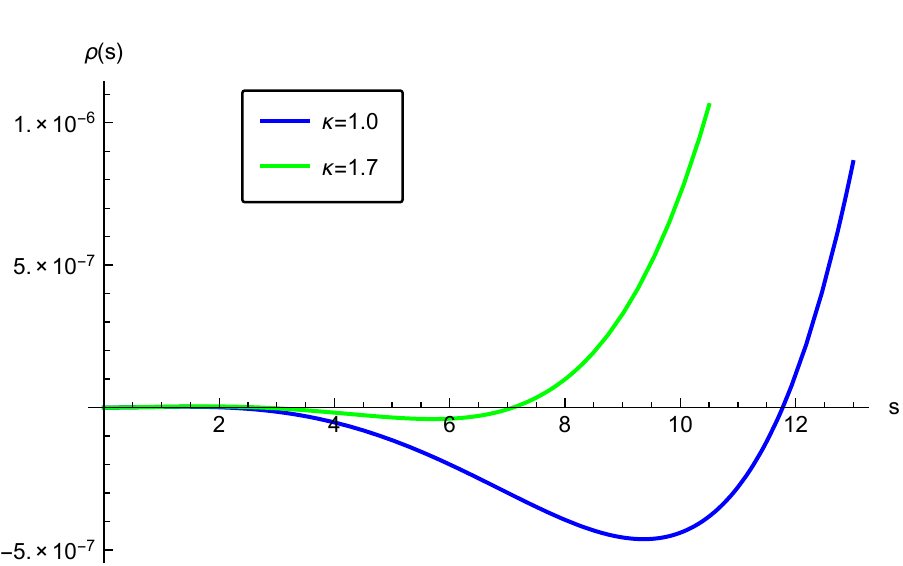}
\caption{Behavior of the spectral density $\rho(s)$ for the scalar current $J^{\Omega\Omega}(x)$.}
\label{fig:sdscalar}
\end{figure}

Besides the tensor current, we shall also investigate the $\Omega\Omega$ diabryon with $J^P=0^+$ by using the 
scalar interpolating current $J^{\Omega\Omega}(x)$ in Eq.~\eqref{scalarcurrent}. Before performing the mass sum 
rule analysis, we study the spectral density and find that $\rho(s)$ is negative in a broad region 
2 GeV$^2$ $\leq s\leq 12$ GeV$^2$, as shown in Fig.~\ref{fig:sdscalar} ($\kappa=1.0$). Such behavior of the spectral density is distinct from the situations of the tensor current for all $J^P=0^+$ and $2^+$ channels, in which the negative regions of the spectral densities are rather small. In the case of the scalar current, the negative quark-gluon mixed condensate $m_s\dqGq$ gives more contribution to the spectral density than those in the tensor current. To eliminate such negative effect, we consider the violation 
of factorization assumption by varying the four-quark condensate $\langle\bar s\bar sss\rangle=\kappa\dqq^2$~\cite{QCD-sum-rule-Reinders}. As shown Fig.~\ref{fig:sdscalar}, the behavior of spectral density is good enough for $\kappa=1.7$ of the factorization assumption. We will use such modified spectral density in our mass sum rule analysis of the scalar current. 

\begin{figure}[ht]
\includegraphics[width=0.43\textwidth]{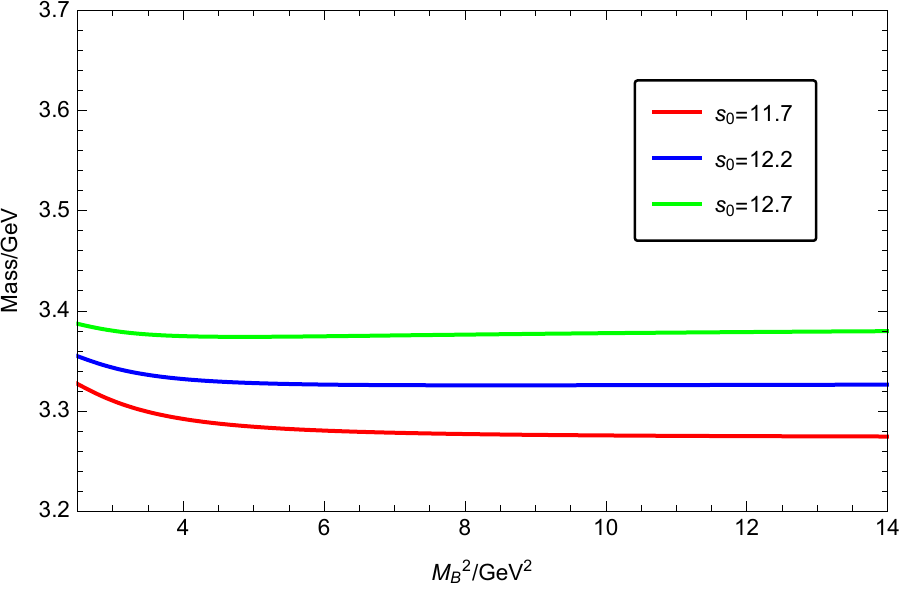}
\caption{Borel curves of the hadron mass depending on $\Msq$ for the scalar current $J^{\Omega\Omega}(x)$.}
\label{fig:massscalar}
\end{figure}

According to the same criteria for $s_0$ and $M_B$ as above, we choose parameters 11.5 GeV$^2$ $\leq s_0\leq 12.7$ GeV$^2$ and 5.0 GeV$^2$ $\leq M_B^2\leq 12$ GeV$^2$. The Borel curves in Fig.~\ref{fig:massscalar} show very good stability and the dibaryon mass is extracted as 
\begin{align}
m_{\Omega\Omega}=(3.33\pm0.42)\, \mbox{GeV}\, , \label{massscalar}
\end{align}
with the coupling constant  
\begin{align}
f_{\Omega\Omega}=\left( 2.1\pm0.16 \right) \times 10^{-3} \, \unit^8\, .
\end{align}
%

Cosidering both Eq.~\eqref{massTtensor} and Eq.~\eqref{massscalar}, the mass of the scalar $\Omega\Omega$ dibaryon state with $J^P=0^+$ can be predicted as 
\begin{align}
m_{\Omega\Omega,\, 0^+}=(3.33\pm0.22)\, \mbox{GeV}\, . \label{Final:scalarmass}
\end{align}
This value is about 15 MeV below the $2m_{\Omega}$ threshold~\cite{PDG}, suggesting the existence of a loosely bound molecular state of the scalar $\Omega\Omega$ dibaryon. Within errors, our prediction of the binding energy is in good agreement with the recent HAL QCD result~\cite{2018-Gongyo-p212001-212001}, while much smaller than the chiral SU(3) quark model calculation~\cite{2000-Zhang-p65204-65204}. 

\begin{figure}[ht]
	\includegraphics[width=0.43\textwidth]{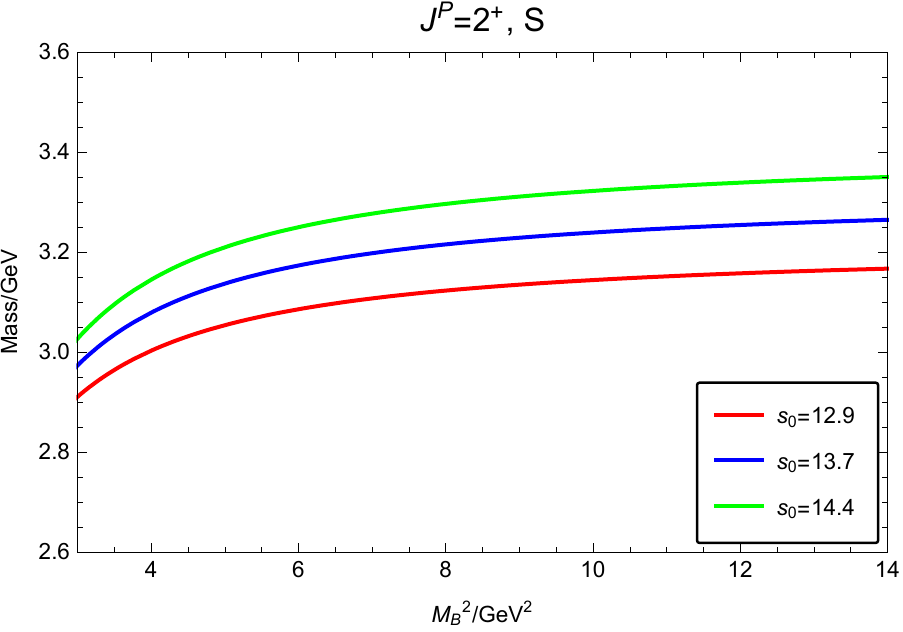}
	\caption{Borel curves of the extracted mass for the $J^P=2^+$ dibaryon using the tensor current $J^{\Omega\Omega}_{\mu\nu}(x)$.}
	\label{fig: result tensor 2}
\end{figure}

{\it Prediction for the tensor $\Omega\Omega$ Dibaryon with $J^P=2^+$:}
To investigate the tensor $\Omega\Omega$ dibaryon state, we use the projector $P_{2S}^P$ 
in Eq.~\eqref{projectors} to pick out the tensor invariant structure in $\Pi_{\mu\nu,\,\rho\sigma}(q^2)$. 
Using this invariant function, we perform similar analysis as for the scalar channel. 
Unfortunately, we find that the Borel curves do not stabilize for any value of $s_0$ for the tensor channel, 
as shown in Fig.~\ref{fig: result tensor 2}. The reason for that is the sign reverse of the quark condensate 
$m_s\dqq$ contribution, which is the dominant non-perturbative term in the OPE series. 

In order to make a mass prediction, we need to fix the value of $s_0$ under an additional criterion. 
The mass splitting between the tensor and scalar $\Omega\Omega$ dibaryons is mainly due to the spin-spin interaction of the two $\Omega$ baryons. This is similar with the mass difference between the non-strange dibaryons $D_{10}$ and $D_{12}$ in SU(6) symmetry, where $m_{D_{12}}-m_{D_{10}}\approx284$ MeV~\mycite{dibaryon-sextet,dibaryon-review}. 
For the heavier $\Omega\Omega$ systems, it is thus reasonable to fix the continuum threshold by assuming 
$\sqrt{s_{0, 2^+}}-\sqrt{s_{0, 0^+}}\approx$0.1-0.3 GeV. This criterion leads to the working region 12.9 GeV$^2$ $\leq s_0\leq 14.4$ GeV$^2$ (central value $s_0=13.7$ GeV$^2$) in the tensor sum rule analysis. 
We plot the mass curves depending on $M_B^2$ in Fig.~\ref{fig: result tensor 2}. One can trust this approximation providing a reliable mass prediction to the tensor $\Omega\Omega$ dibaryon with $J^P=2^+$
\begin{align}
m_{\Omega\Omega,\, 2^+}=(3.24\pm0.23)\, \mbox{GeV}\, , \label{masstensor}
\end{align}
with the coupling constant  
\begin{align}
f_{\Omega\Omega,\, 2^+}=\left( 1.8\pm 0.7 \right) \times 10^{-3} \, \unit^8\, .
\end{align}
The above predicted mass is about 100 MeV below the $2m_\Omega$ threshold, which may imply a deeper molecular state of the tensor 
$\Omega\Omega$ dibaryon than the scalar channle. This result is different from the weakly repulsive interaction for the tensor $\Omega\Omega$ system obtained by the lattice QCD calculation at a pion mass of 390 MeV in Ref.~\cite{2012-Buchoff-p94511-94511}.
	
{\it Summary and Discussion:}
In this work, we have investigated the scalar and tensor $\Omega\Omega$ dibaryon states in $^1S_0$ and 
$^5S_2$ channels with $J^P=0^+$ and $2^+$ respectively in the framework of QCD sum rules. We construct the scalar and tensor $\Omega\Omega$ dibaryon interpolating currents in a molecular picture, by which the spectral densities and two-point correlation functions are calculated up to dimension ten condensates. 

We use different 
projectors to pick out spin-0 and spin-2 invariant structures from the tensor correlation function, and find that all the 
trace part (T), the traceless symmetric part (S) and the cross term (TS) couple to the $0^+$ dibaryon. However, 
the numerical analyses show that only the trace part from the tensor current and the scalar current can provide 
stable mass sum rules. Accordingly, we make reliable mass prediction for the scalar $\Omega\Omega$ dibaryon 
with $J^P=0^+$ to be $m_{\Omega\Omega,\, 0^+}=(3.33\pm0.22)\, \mbox{GeV}$. This value suggests the existence 
of a loosely bound scalar $\Omega\Omega$ dibaryon with a small binding energy around 15 MeV. Our result supports the attractive interaction existing in the scalar $\Omega\Omega$ channel, with the small binding energy in agreement with the HAL QCD simulation~\cite{2018-Gongyo-p212001-212001}. For the tensor $\Omega\Omega$ system, our result provides a mass 
prediction around $m_{\Omega\Omega,\, 2^+}=(3.24\pm0.23)\, \mbox{GeV}$, which may imply a deeper molecular 
state of the tensor $\Omega\Omega$ dibaryon than the scalar channel. 

Since their strangeness $S=-6$ and masses are below the two-$\Omega$ threshold, the $\Omega\Omega$ dibaryons can 
only decay under the weak interaction. In such a molecular system, an $\Omega$ component in the $\Omega\Omega$ dibaryon 
can decay as a free particle while another $\Omega$ acting as the spectator throughout the whole process. Therefore the dominant 
decay modes for the scalar $\Omega\Omega$ dibaryon are $\Omega\Omega\to\Omega^-+\Lambda+K^-$, $\Omega\Omega\to\Omega^-+\Xi^0+\pi^-$ and $\Omega\Omega\to\Omega^-+\Xi^-+\pi^0$, while only the latter two processes exist for the tensor channel. Moreover, both of the scalar and tensor $\Omega\Omega$ dibaryons may decay into the $\Xi\Xi K$ final states. Such exotic strangeness $S=-6$ and doubly-charged $\Omega\Omega$ dibaryon states may be produced and identified in the heavy-ion collision experiments in the future.

\section*{ACKNOWLEDGMENTS}

We thank Prof. Shi-Lin Zhu for useful discussions. This project is supported by the National Natural Science Foundation of China under Grants No. 11722540, the Fundamental Research Funds for the Central Universities.


\begin{widetext}
\newpage 
\appendix*
\section{Supplemental Material}
We collect here the expressions of spectral densities for both the scalar and tensor interpolating currents $J^{\Omega\Omega}(x)$ and $J^{\Omega\Omega}_{\mu\nu}(x)$ described in the text. 

\begin{itemize}
\item For the scalar current with $J^P=0^+$:
\begin{equation}
\begin{split}
	\rho(s) &=
	\frac{27 s^7 }{ 7! 7! 2^9 \pi^{10}} 
	-\frac{21 m_s \dqq s^5}{5! 5! 2^5 \pi^8} 
	-\frac{9 \dGG s^5}{5! 5! 2^{11} \pi^{10}} 
	-\frac{3 m_s \dqGq s^4}{5! 2^{5} \pi^8}
	+\frac{17 \dqq^2 s^4}{5! 2^6 \pi^6}   
	+\frac{11 m_s \dGG \dqq s^3}{4! 4! 2^{4} \pi^8} 
	+\frac{3 \dqGq \dqq s^3}{2^7 \pi^6}
	\\
			&\quad 
	+\frac{5 m_s \dGG \dqGq s^2}{2^{10} \pi^8} 														
	-\frac{17 m_s \dqq^3 s^2}{12 \pi^4} 
	+\frac{13 \dqGq^2 s^2}{2^8 \pi^6}
	-\frac{221 \dGG \dqq^2 s^2}{4! 5! 2^2 \pi^6}\, .
\end{split}
\end{equation}

\item For the trace part of the tensor current ($J^P=0^+$, T) 
\begin{equation}
\begin{split}
	\rho(s) &= 
	\frac{27 s^7 }{ 7! 7! 2^{13} \pi^{10}} 
	-\frac{7 m_s \dqq s^5}{5 \times 5! \times 2^{12} \pi^8} 
	-\frac{31 \dGG s^5}{5^2 2^{27} \pi^{10}} 
	-\frac{3877 m_s \dqGq s^4}{5! 2^{20} \pi^8}
	+\frac{\dqq^2 s^4}{3 \times 2^{12} \pi^6}   
	+\frac{491 m_s \dGG \dqq s^3}{3^3 2^{21} \pi^8} 
	\\
			&\quad 
	+\frac{4645 \dqGq \dqq s^3}{3^2 2^{19} \pi^6}+\frac{865 m_s \dGG \dqGq s^2}{9 \times 2^{20} \pi^8} 
	-\frac{5 m_s \dqq^3 s^2}{3 \times 2^5 \pi^4} 
	+\frac{2725 \dqGq^2 s^2}{3 \times 2^{19} \pi^6}												  
	-\frac{23 \dGG \dqq^2 s^2}{3^3 2^{14} \pi^6}\, .
\end{split}
\end{equation}

\item For the traceless part of the tensor current ($J^P=0^+$, S) 
\begin{equation}
\begin{split}
	\rho(s) &= 
	\frac{3 s^7 }{ 7! 7! 2^{12} \pi^{10}} 
	-\frac{57 m_s \dqq s^5}{5 \times 7! 2^{11} \pi^8} 
	+\frac{2693 \dGG s^5}{15 \times 7! 2^{24} \pi^{10}} 
	 -\frac{46073 m_s \dqGq s^4}{3 \times 7! 2^{19} \pi^8} 
	+\frac{3 \dqq^2 s^4}{7! 2^5 \pi^6}	
	+\frac{2653 m_s \dGG \dqq s^3}{5 \times 3^4 2^{18} \pi^8} 
	\\
			&\quad 
	+\frac{3011 \dqGq \dqq s^3}{5 \times 3^3 2^{19} \pi^6}
	+\frac{2083  m_s \dGG \dqGq s^2}{3^3 2^{20} \pi^8} 
	-\frac{5 m_s \dqq^3 s^2}{3^2 2^3 \pi^4} 
	-\frac{269 \dqGq^2 s^2}{3 \times 2^{18} \pi^6}												  
	-\frac{317 \dGG \dqq^2 s^2}{3^3 2^{16} \pi^6}\, .
\end{split}
\end{equation}

\item For the cross term of the tensor current ($J^P=0^+$, ST) 
\begin{equation}
\begin{split}
	\rho(s) &= 
	\frac{m_s \dqq s^5}{5 \times 7! \times 2^{12} \pi^8} 
	+\frac{367 \dGG s^5}{7! 2^{23} \pi^{10}} 
	+\frac{18389 m_s \dqGq s^4}{5 \times 3^3 2^{21} \pi^8}
	-\frac{\dqq^2 s^4}{5! 2^{6} \pi^6}   
	-\frac{1183 m_s \dGG \dqq s^3}{45 \times 2^{21} \pi^8} 
	-\frac{18389 \dqGq \dqq s^3}{45 \times 2^{18} \pi^6}
	\\
			&\quad 
	-\frac{5 \dGG \dqq^2 s^2}{3^3 2^{17} \pi^6}
	-\frac{1927 m_s \dGG \dqGq s^2}{3^2 2^{21} \pi^8} 
	+\frac{m_s \dqq^3 s^2}{16 \pi^4} 
	-\frac{11477 \dqGq^2 s^2}{3^2 2^{19} \pi^6}\, .
\end{split}
\end{equation}

\item For the tensor current ($J^P=2^+$, S) 
\begin{equation}
\begin{split}
	\rho(s) &= 
	\frac{3 s^7 }{ 7! 7! 2^7 \pi^{10}} 
	+\frac{47 m_s \dqq s^5}{7! 2^9 \pi^8} 
	-\frac{188399 \dGG s^5}{9 \times 7! 2^{22} \pi^{10}} 
	-\frac{25 \dqq^2 s^4}{7! 24 \pi^6} 
	+\frac{822245 m_s \dqGq s^4}{9 \times 7! 2^{16} \pi^8}
	+\frac{6149 m_s \dGG \dqq s^3}{3^5 2^{18} \pi^8} 
	\\
			&\quad 
	-\frac{16823 \dqGq \dqq s^3}{3^4 2^{15} \pi^6}
	-\frac{18401 \dqGq^2 s^2}{3^2 2^{17} \pi^6}	
	-\frac{25 m_s \dqq^3 s^2}{3^3 2^2 \pi^4} 										  
	-\frac{385 \dGG \dqq^2 s^2}{3^4 2^{11} \pi^6}
	+\frac{3671 m_s \dGG \dqGq s^2}{3^4 2^{17} \pi^8}\, .
\end{split}
\end{equation}

\end{itemize}
\end{widetext}

\end{document}